\documentclass[12pt]{article}                            
\usepackage{graphicx}
\newcommand{\comma}{\;\;\; ,}
\newcommand{\period}{\;\;\; .}
\newcommand{\eq}{\, = \,}
\newcommand{\sep}{\,\; , \;\;}
\newcommand{\be}{\begin{equation}}
\newcommand{\bd}{\begin{displaymath}}
\newcommand{\ee}{\end{equation}}
\newcommand{\ed}{\end{displaymath}}
\newcommand{\ba}{\begin{eqnarray}}
\newcommand{\ea}{\end{eqnarray}}
\newcommand{\Kay}{{\cal K}}

\newcommand{\half}{\textstyle \frac{1}{2}}
\newcommand{\third}{\textstyle \frac{1}{3}}
\newcommand{\schoose}{\textstyle \left(
 \begin{array}{c} {\textstyle \!\!\! i+j \! \! \!} \\ {\textstyle\! 3 \! } \end{array}  \right) }

\def\picture #1 by #2 (#3){
  \vbox to #2{
    \hrule width #1 height 0pt depth 0pt
    \vfill
    \special{picture #3}}}
\def\scaledpicture #1 by #2 (#3 scaled #4){{
  \dimen0=#1 \dimen1=#2
  \divide\dimen0 by 1000 \multiply\dimen0 by #4
  \divide\dimen1 by 1000 \multiply\dimen1 by #4
  \picture \dimen0 by \dimen1 (#3 scaled #4)}}

\title{Dichromatic polynomials and Potts models summed over rooted maps}

\author{ R.J. Baxter\\
{\protect \small School of Mathematical
Sciences  and  Research School }\\ 
{\protect \small  of Physical Sciences \& Engineering}\\
{\protect \small  The Australian National University,
 Canberra, A.C.T. 0200, Australia  }
}

\date{11 January 2001}

\begin{document}

\maketitle

\abstract{We consider the sum of dichromatic polynomials over non-separable
rooted planar maps, an interesting special case of which is the enumeration of such
maps. We present some known results and derive new ones. The general 
problem  is equivalent to the $q$-state Potts model 
randomized over such maps. Like the regular ferromagnetic lattice models, it has a 
first-order transition when $q$ is greater than a critical value $q_c$, 
but $q_c$ is much larger - about 72 instead of 4.
}

\section*{Introduction}
The dichromatic polynomial of a graph $\cal G$ is a polynomial $\chi ({\cal G};x,y)$ 
in two variables
$x, y$. It has been 
studied by Whitney\cite{Whitney32}, Tutte\cite{Tutte67} and others.
In particular, Tutte obtained a functional relation for the generating function
$\Phi$ of the sum of dichromatic polynomials over rooted planar maps.\cite{Tutte71}

 Fortuin and Kasteleyn\cite{Fortuin-Kasteleyn,Kasteleyn-Fortuin} showed, from the
viewpoint of statistical physics, that $\chi ({\cal G};x, y)$ is 
equivalent to the partition function of a $q$-state Potts model on $\cal G$, 
with $q = (x-1) (y-1)$. This means that
$\Phi$ is the partition function of the sum of maps. From it one should be able to obtain
the thermodynamic properties of the planar Potts model, averaged over all rooted maps.
This is not without interest: usually one considers such models on a regular lattice,
but physical lattices are seldom (if ever) completely regular. They contain defects, and 
one should in fact average over such defects, i.e. over a lattice with some degree 
of randomness. In this sense $\Phi$ gives the average over over a completely random
``lattice''. 

There has also been much work in statistical mechanics and field theory on
summing over planar Feynman diagrams, using field-theoretic techniques and 
matrix models.\cite{'t_Hooft74,Brezin78,Daul95,
Kostov95,Kostov00,Wexler93,Ambjorn95,ZinnJustin99,ZinnJustinZuber99} The main difference between that approach and the one 
used here is that our maps are ``rooted'', which means that they 
are given extra weights, corresponding to the various ways a given map can be rooted.
It appears that this is merely a ``bounday effect''which does not influence the asymptotic 
large-graph behaviour. 

Here we do try to reduce the randomness by restricting each map $\cal G$ to be
non-separable (i.e. irreducible). The required analogue of Tutte's functional relation
for this case has been obtained by Liu (eqn. 4.17 of \cite{Liu90}).  There are three cases for 
which the relation can be solved explicitly:

\begin{tabbing}
Case 1:    \=$x,y$ both large, of the same order; \\ 

Case 2:    \>$x, y$ both small, of the same order; \\

Case 3:    \>$q =  $ \= $ \, (x-1) (y-1) = 1$,~  or $x$ large and $y$ of order one, \\
															\> \> 			or $y$ large and $x$ of order one.
\end{tabbing}

We present the solution in all three cases. The first is rather trivial. The second 
appears to be both non-trivial and new: one interesting feature we observe is
that for  non-separable maps $M$ of two or more edges 
\be  \label{conj}
\chi (M;x, y) \sim c_M (x+y) + {\rm higher \; terms \; in } \; \; x, \, y \comma \ee
the coefficients $c_M$ of $x$ and $y$ being the same positive integer. We  
first verified this on the  computer for 
maps of up to ten edges.  Since then Professor Tutte has pointed 
out to the author that (\ref{conj}) can be proved in general by recursively using 
Theorems 1 and 2 of Appendix A.

 The last case is equivalent (to within a re-definition of $v_1, v_2$) to  
$\chi ({\cal G}) = 1$  for all maps  $\cal G$, i.e. to a
Potts model with no interaction (the ideal lattice gas). In this case the problem 
reduces to the enumeration
of non-separable  rooted planar maps. As the author discovered after solving this case,
there is an extensive literature on this 
subject.\cite{Tutte63,Tutte68,Walsh75,Wormald81,Liskovets83,Bender85,Arques85,
Liu88,Arques99} Explicit results have been obtained by Tutte 
(eq. 6.4 of Ref. \cite{Tutte63}) and Brown and Tutte.\cite{Brown63,BrownTutte64} 
They involve the intriguing 
hypergeometric function identities (\ref{pmqn}), (\ref{pm}).

We also extend case 3 slightly, conjecturing the second-order corrections for $x$ large
when $y = 1$.

In statistical mechanics we are particularly interested in large
systems, when we expect the bulk properties to be obtainable by taking an 
appropriate limit. Combinatorially this corresponds to the asymptotic
behaviour \cite{Bender86} of the sum of dichromatic polynomials. In all our solved cases 
we find that the appropriate limit does indeed exist. 

For the regular planar lattices we know that the $q$-state Potts model has 
a first-order phase transition (i.e. a non-analyticity in the bulk properties) 
for $q > q_c$, and  a continuous transition for 
$1 < q \leq q_c$, where $q_c = 4$. \cite{Baxter73,BaxTempAsh78,Baxter82,Wu82} 
It seems from numerical sudies that the random model of this paper behaves similarly, 
but with $q_c \simeq 72$.

The outline of this paper is as follows. In the next section we define 
the dichromatic polynomial and the Potts model, and state their equivalence.
Then we introduce the concept of non-separable rooted planar maps, and 
of the sum $\Phi$ over their dichromatic polynomials. We give the functional relation
satisfied by $\Phi$.  We then discuss the three explicitly solved cases, and go on to 
consider the asymptotic behaviour in the large-map limit. Finally we discuss 
the  phase transition and numerical methods that can be used to 
study this.

\section*{The dichromatic polynomial and the Potts 
model}

Let $\cal G$ be any connected graph (planar or not). Then the dichromatic polynomial is
defined in \cite{Tutte71} as
\be \label{dichpoly}
\chi ({\cal G};x,y) \eq \sum_S  (x-1)^{C(S)-1} \, (y-1)^{C(S)+E(S)-V(S)}  \comma \ee
where $x,y$ are arbitrary variables, the sum is over all sub-graphs $S$ of $\cal G$, and 
$C(S), E(S), V(S)$ are the numbers of connected components, edges and vertices, respectively, 
of $S$. Since $S$ and $\cal G$ share the same vertices, $V(S) = V({\cal G})$. We have used Euler's relation: the number of circuits of $S$ is 
\be \label{circuits}
P(S) = C(S)+E(S)-V(S) \period \ee
Thus for example the dichromatic polynomial of the graph of two vertices connected by a
single edge is $(x - 1) + (1) = x$; for a triangle of three vertices connected in pairs by three edges
it is $(x-1)^2+3(x-1)+3(1)+(y-1) = x^2+x+y$.

A $q$-state Potts model is defined on $\cal G$ by associating with each vertex $i$  
a ``spin'' $\sigma_i$
that takes the values $1, \ldots , q$. Let $K, v$ be two variables related by
\be v = e^K   - 1 \period \ee 
Then the partition function is
\be \label{ZPotts1}
Z({\cal G};q,v) \eq \sum_{{\bf \sigma}} \exp \{ K \sum_{(i,j)}
 \delta(\sigma_i, \sigma_j ) \} \comma \ee
where the outer sum is over all $q^{V({\cal G})}$ values of all the spins, 
the inner sum is over all edges $(i,j)$ of $\cal G$, and  
$\delta(a,b) = 1$ if $a = b$, $\delta(a,b) = 0$ if $a \neq b$.

The eqn. (\ref{ZPotts1}) can be written as
\be \label{ZPotts2}
Z({\cal G};q,v) \eq \sum_{{\bf \sigma}} \prod_{(i,j)}
 \{ 1 + v \delta(\sigma_i, \sigma_j ) \} \period \ee
There are $E({\cal G})$ factors in the product, one for every edge 
of $\cal G$. Expanding the product, we obtain $2^{E({\cal G})}$ terms, 
each of which can be represented by a sub-graph $S$ of $\cal G$ by
including an edge in $S$ if we take the $ v \delta(\sigma_i, \sigma_j ) \}$ term, 
leaving it out if we take the leading term $1$.

For each of the  $2^{E({\cal G})}$ terms thus obtained, we can readily perform 
the sum over the spins, giving
\be 
Z({\cal G};q,v) \eq \sum_S q^{C(S)} \, v^{E(S)}  \comma \ee
so we see that
\be 
Z({\cal G};q,v) \eq (x-1) \, (y-1)^{V({\cal G})} \, \chi ({\cal G};x,y)  \ee
provided that
\be
q = (x-1)(y-1) \sep v = y-1 \period \ee

\section*{Non-separable rooted planar maps}

\begin{figure}[hbt]            
\centering
\includegraphics[scale=0.6,angle=0]{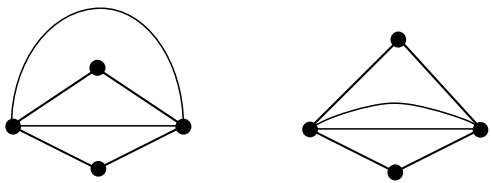}     
\caption{\protect\small Two maps corresponding to the same graph.  }
\label{Two-graphs}
\end{figure}

A planar map (or simply a map) is a connected graph embedded in the surface of a sphere or closed plane, with no crossing 
edges. Not all graphs are planar, and those that are may correspond to 
more than one planar map: for instance the maps in Figure \ref{Two-graphs} are different, but 
correspond to the same graph. If the vertices and faces are given distinctive labels, then
a map can be specified uniquely by listing the anti-clockwise cyclic sequence of alternating 
vertices 
and faces (two of  each) round each edge. Two maps are distinct if and only if 
their edge lists cannot be made the same by re-labelling the vertices and sites, 
and re-ordering the list.


\vspace{6.5mm}                                     
 
A map $M$ is {\em rooted} if one edge $A$ is chosen as the root-edge 
and given a direction. The vertex at the start of this directed edge is known as the
{\em root-vertex} , the face on the left as the {\em root-face}. Two rooted maps 
are distinct if and only if 
their edge lists, root-edge, root-vertex and root-face cannot all be made the same 
by re-labelling the vertices and sites, 
and re-ordering the list.

\begin{figure}[hbt]                    
\centering
\includegraphics[scale=0.6,angle=0]{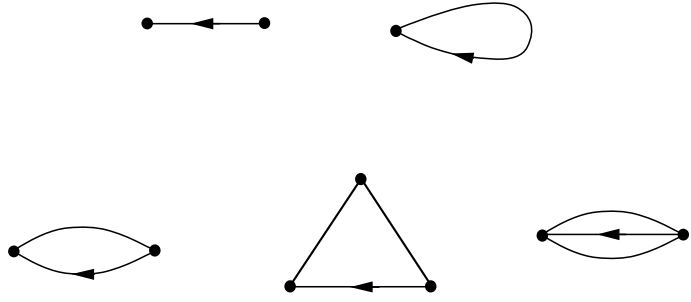}     
\caption{\protect\small  Non-separable rooted maps with up to three edges.}
\label{simple_graphs}
\end{figure}

Finally, a graph is {\em non-separable} if there is no vertex such that the 
graph obtained by 
cutting it at this vertex (i.e. disconnecting from the vertex all 
its incident edges)
is no longer connected.  The non-separable 
rooted maps of up to three edges are shown in Figure \ref{simple_graphs}. Their dichromatic polynomials, 
as defined by (\ref{dichpoly}), are $x, y, x+y, x^2+x+y, x+y+y^2$, respectively. 
The first two are the  {\em link-map} 
and the {\em loop-map}: they are the smallest maps we shall consider. Note 
that we allow repeated edges, but only the loop-map contains a loop, and only the 
link-map has sites of valence one.


\vspace{6.5mm}                                     
 
Here we follow the original notation of Tutte\cite{Tutte71} rather than Liu.\cite{Liu90} For a given map $M$,  let $m, n, i, j$ be the valency of the root-face, the valency of the
 root-vertex, the number of faces other than the root-face, and the number of vertices other 
than the root-vertex. Then the number of edges of $M$ is $i+j$. We say that a map is ``of type 
$(m, n, i, j)$'' if it has these values of $m, n, i, j$. Define
\ba \label{defnPhi}
\Phi & = & \Phi(u_1,u_2)  \eq \Phi(u_1,u_2,v_1,v_2,x,y) \nonumber \\
& = & \sum_M u_1^m u_2^n v_1^i v_2^j \chi (M;x,y) \comma \ea
where the sum is over all distinct rooted non-separable maps $M$ and $m,n,i,j$ are the numbers 
defined above for each map $M$. Writing the contributions from the link and 
loop maps explicitly, this is
\be \Phi(u_1, u_2)  \eq x u_1^2 u_2 v_2 + y u_1 u_2^2 v_1 + \Phi'(u_1,u_2)  \comma \ee
where the sum $\Phi'(u_1,u_2)$ is  further restricted to  maps  of two or more edges.
Such maps also have $i, j \geq 1$ and $m, n \geq 2$,  so 
\be \label{defN}
\Phi' (u_1, u_2 ) \eq \sum_{m, n, i, j} N(m, n, i, j) \, u_1^m u_2^n v_1^i v_2^j  \comma \ee
where
\be \label{ineq} 1 \leq  m-1 \leq j \sep 1 \leq  n - 1 \leq i
\period \ee
The coefficient $N(m, n, i, j)$ is the sum of the dichromatic polynomials over 
all distinct non-separable rooted maps
of type $(m, n, i, j)$. We also define coefficients $N', P, R$ by
\ba \label{defPR}
\Phi' (1, u_2) & = & \sum_{n, i, j} N'(n, i, j) \, u_2^n v_1^i v_2^j  \comma \nonumber \\
\Phi' (1, 1) & = & \sum_{i, j} P( i, j) \, v_1^i v_2^j  \period \\
\Phi' (1, 1,v,v,x,y) & = & \sum_{k=2}^{\infty} R (k)  \, v^k  \period \nonumber \ea
Thus $P(i,j)$ is the sum over all distinct non-separable rooted maps with $i+1$ 
faces and $j+1$ vertices; 
$R(k)$ is the sum over all such  maps with $k$ edges. Clearly
\bd
N'(n,i,j) \eq \sum_{m = 2}^{j+1} N(m, n, i, j)  \comma \ed
\be \label{PRsum} 
P(i,j) = \sum_{n=2}^{i+1} N' (n,i,j) \sep R(k) = \sum_{i=1}^{k-1} P(i,k-i) \ee

From Figure \ref{simple_graphs} and the 
comments above,
\bd \Phi' (u_1, u_2 )  =  (x+y) u_1^2 u_2^2 v_1 v_2 + 
(x^2+x+y) u_1^3 u_2^2 v_1 v_2^2 + (x+y+y^2) u_1^2 u_2^3 v_1^2 v_2 + \cdots \ed
all other terms being of order $v_1^i v_2^j$, with $i + j > 3$.

Tutte shows that the concept of {\em duality} extends to rooted maps. If $M^*$ is the dual of $M$,
and $m^*, n^*, i^*, j^*$ the corresponding values of $m, n, i, j$, then
\be
m^* = n \sep n^* = m \sep i^* = j \sep j^* = i \comma \ee 
and 
\be \chi (M; x,y) \eq \chi (M^*; y,x) \period \ee

The dual of a non-separable map is also non-separable, so
\be \Phi(u_1,u_2,v_1,v_2,x,y) \eq \Phi (u_2,u_1,v_2,v_1,y,x) \period \ee
We can use this duality to obtain the sum over $n$ of $N(m, n, i, j)$ from $N'$.

Define \ba \label{defC}
C(u_1, u_2) & = &  \frac{u_1 \Phi(1,u_2) - \Phi(u_1,u_2) }{1 - u_1 } \comma \nonumber \\
D(u_1, u_2) & = &  \frac{u_2 \Phi(u_1,1) - \Phi(u_1,u_2) }{1 - u_2 } \period \ea

Then Liu\cite{Liu90} shows that $\Phi$ satisfies the functional relation
\ba \label{fnreln}
\Phi(u_1,u_2 ) & \eq & x u_1^2 u_2 v_2 + y u_1 u_2^2 v_1 + \frac{u_1 u_2 v_1 C(u_1,u_2) }
{1-C(u_1,1)} \nonumber \\
& & \;\;\;\;\; + \;  \frac{u_1 u_2 v_2 D(u_1, u_2) }{1-D(1,u_2)  }\period \ea
For completeness we give the derivation of this result in Appendix A.

We can solve this equation iteratively to obtain the contributions to 
$\Phi$ from non-separable rooted maps of
1, 2, 3, \ldots edges. As we shall find in the next section, 
the numbers of such distinct maps are 
2,1,2,6, 22, 91, 408, 1938, 9614, 49335,....
(e.g. there are 91 non-separable rooted maps with six edges).

As a check, we have enumerated all non-separable maps 
of up to 10 edges, and their various rootings and  
dichromatic polynomials, directly on the computer. 
We have verified that the results 
for $\Phi$ do in fact agree with those obtained from  (\ref{fnreln}). We also find
that the numbers of distinct unrooted maps are 
2,1,2,3,6, 16, 42, 151, 596, 2605,... 
The problem of counting unrooted maps has quite an extensive literature. 
Wormald \cite{Wormald81}, and Liskovets and Walsh \cite{Liskovets83},
established linear relations between the numbers of unrooted and rooted maps: both
the above sequences of integers appear in Figure 1 of Ref. \cite{Liskovets83}.
The relation between them is given in Appendix B.

\section*{Case 1: $x, y$ both large}

One case that can be easily handled is when $x, y$ are 
both large and of the same order. In this case the coefficients in $\Phi$ of $v_1^i v_2^j$, 
for  $i,j \geq 1$ and  arbitrary $u_1, u_2$, are of order $i+j+\delta_{i1}+\delta_{j1}-2$.
To obtain them it is sufficient to set $v_1 = w_1/y, v_2 = w_2/x$, 
take $w_1, w_2$ to be  fixed
and of order one, and then to solve (\ref{fnreln}) iteratively in 
inverse powers of $x$ and $y$. To first order the the result is
\bd
\! \!   \Phi (u_1, u_2)  =  u_1^2 u_2 w_2 \! + \! u_1 u_2^2 w_1 + y^{-1} \frac{u_1^2 u_2^2 w_1 w_2}{1-u_1 w_2} 
 + x^{-1} \frac{u_1^2 u_2^2 w_1 w_2}{1-u_2 w_1} + \cdots \ed
Continuing to second order, then setting $u_1 = u_2 = 1$, we obtain
\ba & & \! \! \! \! \! \!  \Phi(1,1) \eq w_1 + w_2 + \mu_1 w_1 w_2/x + 
w_1 \mu_2 w_2 /y + \mu_1^4 w_1 w_2^2/x^2 +
 \nonumber \\
& &  \mu_1^2 w_1 \mu_2 w_2^2/(x y) +\mu_1 w_1^2 \mu_2^2 w_2/(x y) + w_1^2 \mu_2^4 w_2/y^2 + \cdots
  \ea
where for brevity we have written $1/(1-w_i)$ as $\mu_i$.
From this and (\ref{defPR}) it follows that if $i = 1$ or $j = 1$
\bd  P(i, j) \sim x^j \delta_{i \, 1} +y^i \delta_{j \, 1} \comma \ed
if $i, j \geq 2 $
\be P(i, j) \sim (i+j) \, x^{j-1} y^{i-1}  + \schoose
(x^j \delta_{i \, 2} +y^i \delta_{j \, 2} ) \comma \ee
 and
\be R(k) \sim  x^{k-1} + y^{k-1}  \period \ee

\section*{Case 2: $x, y $ small}

For any graph, the dichromatic polynomial vanishes when
$x = y = 0$.
Consider  the limit when $x, y$ are small but non-zero, of the same order. Then any 
dichromatic polynomial is at most of this order. From (\ref{defnPhi}) and (\ref{defC}), 
 so are 
$\Phi, C$ and $D$. If we define $Q = Q(u_1,u_2) = Q(u_1, u_2, v_1, v_2)$
by
\be \label{defQ}
\Phi (u_1 , u_2 ) \eq x \, u_1^2 u_2 v_2 +  y \, u_1 u_2^2 v_1 +
 (x+y) u_1^2 u_2^2 v_1 v_2 \, Q(u_1 , u_2 ) 
\comma \ee
then we find from (\ref{fnreln}) that $x, y$ cancel out of the functional relation  
for $Q$, leaving
\ba \label{Qeqn}
Q(u_1, u_2) &  = & 1 + u_2 v_1 \; \frac{u_1 Q(u_1,u_2) \! - \! Q(1,u_2)}
{ u_1 - 1} +   \nonumber \\
&& u_1 v_2 \; \frac{u_2 Q(u_1,u_2) \! - \! Q(u_1,1)}
{ u_2 - 1}  \period \ea

This implies that, apart from the contributions of the link-map and the loop-map,
$\Phi (u_1 , u_2, v_1, v_2 )$ is proportional to $x + y$ when $x$ and $y$ are both small. 
As we state in the introduction, the stronger statement (\ref{conj}) is true, for 
{\em every} non-separable map $M$ of 
more than one edge. For all maps of two to five edges $c_M = 1$, for six 
or seven edges it is 1 or 2, for eight to ten edges it is no bigger than the number of 
edges minus five (i.e. 3, 4, 5, respectively). The first map to have $c_M > 1$ is the 
tetrahedron (which is planar).

The relation (\ref{Qeqn}) is a linear equation for $Q(u_1, u_2)$. If we regard 
$Q(1,u_2)$ and $Q(u_1,1)$ as known, we can solve it for $Q(u_1, u_2)$. The result 
is a rational expression with a denominator that is biquadratic in $u_1$, $u_2$.
If $v_1$, $v_2$ are both small and $u_2$ is of order one, the two zeros are when
$u_1 \simeq 1$ and when $u_1 \simeq 1/(u_2 v_2)$.

Suppose $v_1, v_2$ are both small, of order $v$. From (\ref{ineq} ),
 $m \leq i+j$ (for $i+j \geq 2$), so
the coefficient of $u_1^m$ in the expansion of $\Phi$ is of order $v^m$ or smaller.
It follows that the zero near $u_1 = 1$ must also be a zero of the numerator, 
since otherwise it would be a pole of $\Phi$ and the coefficient of $u_1^m$ would 
be of order unity.

This
gives a linear relation between $Q(1,u_2)$ and $Q(u_1,1)$ at this value of $u_1$.
We have not yet found a direct way 
to solve this relation, but we have guessed explicit expressions for the coefficients
in the expansion of $Q(u_1, u_2)$. In terms of the coefficients $N, N', P$ of (\ref{defN}),
(\ref{defPR}), they are:
\bd N(m,n,i,j)  = (x+y) (m-1) (n-1) \lambda (m,n,i,j) \times \ed
\bd \frac{(i+j-2)! \, (i+j-m-1)! \, (i+j-n-1)!}{(i-1)! \,  i! \, (i-n+1)! 
\, (j-1)! \, j! \, (j-m+1)! } \comma \ed
\bd N'(n,i,j) = \frac{(x+y) (n-1) n (i+j-2)! \, (i+j-1)! \,(i+j-n-1)!}
{i! \,  (i+1)! \, (i-n+1)! 
\, (j-2)! \, (j-1)! \, j! } \comma \ed
\be P(i,j) = \frac{2 (x+y) (i+j-2)! \, (i+j-1)! \, (i+j)!}
{(i-1)! \,i! \, (i+1)! \,  (j-1)! \, j! \, (j+1)! } \comma \ee
where
\bd 
 \lambda (m,n,i,j) \eq i m + j n - m n + m + n - 2 i -2 j  \period \ed
These formulae hold for all $m,n,i,j$ satisfying (\ref{ineq}) and $i, j \geq 2$.
The last, for $P(i,j)$, extends to $i, j \geq 1$, while the first two should 
be supplemented with
\bd
N(m,n,i,1)  =  (x+y) \delta_{m 2} \, \delta_{n, i+1} \sep
N(m,n,1,j)  =  (x+y) \delta_{n 2} \, \delta_{m, j+1} \comma  \ed
\bd N'(n,1,j)  =  (x+y) \delta_{n 2} \sep N'(n,i,1)  =  (x+y) \delta_{n,i+1}
 \comma \ed
for $i, j \geq 1$.


It is straightforward to prove that these guesses are correct by
direct substitution into (\ref{defQ}) and (\ref{Qeqn}). One needs 
the symmetry $N(m,n,i,j) = N(n,m,j,i)$ and the identity
\be \label{sumN}
\sum_{k=m}^{j+1} N(k,n,i,j ) \eq \frac{(i+j-m) B(m,n,i,j) N(m,n,i,j) }
{i (i+1) (m-1)  \lambda(m,n,i,j) } \comma \ee
where 
\bd  B(m,n,i,j) = (m-2)(m-1)i (i-n+1) + (j-1) n (i m -m -i +j +1)  \period \ed
The identity (\ref{sumN}) can be verified by taking the 
difference between its $m$ and $m-1$ cases.


\section*{Case 3: enumeration of rooted non-separable maps}

If \be \label{restrict}
q = (x-1)(y-1) = 1 \period \ee 
Then the dependence on $C(S)$ in (\ref{dichpoly})
disappears and we immediately obtain (for all graphs)
\be  \chi({\cal G}; x,y) =
y^E({\cal G})/(y-1)^{V({\cal G}) +1} \period \ee
For our maps $M$, $E(M) = i+j$ and $V(M) = j+1$, so
\be \chi (M; x,y) = x^j y^i \period \ee 
Hence $x$ and $y$ can be absorbed into $v_2$ and $v_1$
and
\be \Phi' (u_1, u_2, v_1, v_2, x, y) \eq T(u_1, u_2, y v_1,  x v_2 ) \comma \ee
where
\be \label{defphihat}
 T(u_1, u_2, v_1, v_2) = T(u_1,u_2) =  \sum_M u_1^m u_2^n v_1^i v_2^j \period \ee
Thus the coefficient in $T(u_1,u_2)$ of $u_1^m u_2^n v_1^i v_2^j$ is simply the number of 
rooted non-separable maps 
of type $(m, n, i, j)$.

This case was solved originally by Brown and Tutte \cite{Brown63,BrownTutte64}, and has 
been further investigated by a number of authors, including Arques \cite{Arques85, Arques99},
Bender \cite{Bender85, Bender86}, Liskovets and Walsh 
\cite{Walsh75,Liskovets83}, Liu \cite{Liu88, Liu90} and Wormald \cite{Wormald81}.

As a combinatoric problem this case is of interest in itself. From the point of view of 
statistical mechanics 
it is very simple, corresponding to the one-state Potts model or (as we show below)
a q-state model with no interactions, i.e. the perfect gas. Even so, we hope
to use it as the first step in examining the interacting Potts model of our random model, 
and we 
certainly do need to understand it if we are to use it as a starting point for 
general values of $x$ and $y$. We therefore present the results here and outline 
the working in Appendix C. It involves some rather surprising identities.

C and D simplify in the same way as $\Phi'$: 
$ C (u_1, u_2, v_1, v_2, x, y) \eq {\hat C }(u_1, u_2, y v_1,  x v_2 )$ and 
 $ D (u_1, u_2, v_1, v_2, x, y) \eq {\hat D }(u_1, u_2, y v_1,  x v_2 )$. 
The definitions (\ref{defC}) become 
\ba \label{hatC}
{\hat C}(u_1, u_2, v_1, v_2) & \! \! \! = & u_1 u_2 v_2 + \frac{ u_1  
T(1, u_2, v_1, v_2) - T(u_1, u_2, v_1, v_2) }{1-u_1} 
\; , \nonumber  \\
& & \\
{\hat D }(u_1, u_2, v_1, v_2) & \! \! \! = & u_1 u_2 v_1 + \frac{ u_2 
 T(u_1, 1, v_1, v_2) - T(u_1, u_2, v_1, v_2) }{1-u_2} 
\; , \nonumber \ea
Replacing $v_1, v_2$ in (\ref{fnreln}) by $v_1/y, v_2/x$ and using these functions 
$T, {\hat C}, {\hat D}$, we find that $x, y$ cancel 
out, except only that
the last two terms (involving $C$ and $D$, respectively) acquire factors $1/y, 1/x$. 
This is the only place $x$ and $y$ now appear. 
Since the relation must hold for all $x, y$
satisfying (\ref{restrict}), and this can be written as $1/x + 1/y = 1$, we 
obtain two relations from (\ref{fnreln}), namely
\be \label{fnreln2}
T(u_1,u_2) = 
\frac{u_1 u_2 v_1 {\hat C}(u_1,u_2) }
{1- {\hat C} (u_1,1)} \eq  \frac{u_1 u_2 v_2 {\hat D}(u_1, u_2) }
{1- {\hat D}(1,u_2)  }\period \ee

Other ways to reduce the dichromatic problem to simple enumeration are to allow $x$ or $y$ to
tend to infinity, keeping the other fixed. For instance, when $x$ is large the sum in (\ref{dichpoly}) is dominated
by sub-graphs $S$ where $C(S)$ attains its maximum value $V(S)$. For all but the loop-map this means the sub-graph with
no edges, so $\chi(M; x,y) \sim x^j$ and we can absorb $x$ into $v_2$. We obtain  
(\ref{fnreln2}), but without the last equality involving ${\hat D}$. 
The resulting 
relation is still sufficient to iteratively determine $T(u_1,u_2)$, and hence its desired
power series coefficients, but it is not
as full a description of the properties.

These equations are fully solved in parametric form in Appendix C. In particular, we
introduce two new variables $p, r $, 
defined in terms of $v_1, v_2$ by
\be \label{v1v2}
v_1 = p (1-r)^2 \sep v_2 = r (1-p)^2 \period \ee
Then
\be \label{T11}
  T(1, 1) = p r (1-p-r) \period \ee
These equations  are given in  section 2 of \cite{BrownTutte64}, our $v_1, v_2, p, r$ being
the $x,y,u,v$ therein. As is noted in section 4 
therein,  products of powers of $p$ and $r$ (and hence of $1-p$ and $1-r$) have
remarkably simple expansions in powers of $v_1$ and $v_2$. Let us 
introduce the standard notation
\ba
 (m)_k & = & m (m+1) (m+2) \cdots (m+k-1) \; \; {\rm for } \; \; k >0 \comma \nonumber \\ 
& = & 1 \; \; {\rm for } \; \; k  = 0 \comma\\
& = & 1/[ (m+k) (m+k+1) \cdots (m-1)] \; \; {\rm for } \; \; k <  0 \comma \nonumber \ea
where $k$ is an integer.

For positive integers $i,j$, define
\be \label{Bmnij}
{\beta}(m,n|i,j) = \frac{4 (n i + m j + m n ) 
(2 i + 2 m + 1)_{j-1} (2 j + 2 n + 1)_{i-1} }{ i! \, j! } 
 \; . \ee
This can be extended to $i$ or $j$ zero by taking an appropriate limit, giving
${\beta}(m,n|0,j) = 2 m  (2 m +1)_{j-1} / j! $, 
${\beta}(m,n | i,0) =  2 n (2 n+1)_{i-1} /i! $, and  ${\beta}(m,n | 0,0 ) = 1 $.
Thus ${\beta} ( m,n | i,j)$ is a polynomial in $m$ and $n$, for 
all non-negative integers $i, j$.

Then we find the following expansion in powers of $v_1$ , $v_2$:
\be \label{pmqn}
p^m r^n \eq \sum_{i = 0}^{\infty} \sum_{j = 0}^{\infty} 
{\beta}(m,n | i,j) v_1^{i+m} v_2^{j+n} \period \ee
This elegant formula is true for all integers $m,n$, positive or negative,
and can be extended to real and complex values of $m, n$.

If $v_1 = v_2 = v$, then $r = p$ and
\be \label{pm}
p^m \eq \sum_{k=0}^{\infty} \alpha(m,k) v^{k+m} \comma \ee
where
\be \alpha(m,k) = 2 m (2 k + 2 m +1)^{k-1} /k! \period \ee
This is also an intriguing result.

The coefficients $P(i,j), R(k)$ in (\ref{defPR}) are now easily determined:
\be \label{Pij3}
P(i,j) = 4  y^i x^j \frac{ 4 \,  (i+2 j-2)! \, (2 i +j -2)! }{
(i-1)! \, (2 i)! \, (j-1)! \, (2 j)! } \comma \ee
and if $x = y$,
\be \label{Rk}
R(k) = x^k \frac{2 (3 k-3)!}{ k! \, (2 k-1)!} \period \ee
Thus $P(i,j)/(y^i x^j)$ is the number of non-separable rooted maps with
$i+1$ faces and $j+1$ vertices, while $R(k)/x^k$ is the number with $k$ edges.
These formulae were first given by Tutte and Brown \cite{Tutte63,BrownTutte64}.

We have not found an explicit formula for the coefficients $N'(n,i,j)$, $N(m,n,i,j)$,
but it does follow from Appendix C that they have a certain structure. Let 
$m' = m-1+\delta_{m 2} $, $n' = n-1+\delta_{n 2} $, $m''= m-\delta_{m 2}-2$ and
$n''= n -\delta_{n 2}-2$. Then
\bd 
N(m, n, i, j) \eq \frac{4 y^i x^j \, (2 i -1)_{j-2 m - n''+2} \;  (2 j-1)_{i-m''-2 n + 2} \, 
G_{m \, n} (i,j)}
{ (i-n')! \, (j-m')! }  \ed
\be \label{N3} 
N'(n, i, j) \eq \frac{4 y^i x^j  \, (2 i +1)_{j-n-1} \; (2 j-1)_{i-2n+2}  \, G'_n (i,j)}
{ (i-n+1)! \, (j-2)! } \period \ee
Here $G_{m \, n} (i,j)$ and $G'_n (i,j)$ are polynomials in $i, j$ of total degree
$2m''+ 2n''$, $2 n - 4$, respectively. By ``of total degree $d$'', we mean that the 
sum of the powers 
of $i$ and $j$ in any term does not exceed $d$. In particular, 
$G_{2 2}(i,j) = 0$, $G_{2 3}(i,j) = 1$, $G_{2,4} (i,j) =
3 (21-15 i + 2 i^2 - 17 j + 5 i j + 4 j^2)$, $G_{3,3}(i,j)$ is of total degree 4,
$G'_2 (i,j) = 1$ and $G'_3 (i,j) = 6 i^2 + 3 i (5 j-11) + 12 (j-1)(j-2)$.

The coefficients $N(m, n, i, j)$, $N'(n, i, j)$ are always finite. For small values 
of $i, j$ the factors in (\ref{N3}) other than $G_{m \, n} (i,j), G'_n (i,j)$ may 
become infinite, but in this case $G_{m \, n} (i,j)$ or $G'_n (i,j)$ vanishes and 
the coefficient can be evaluated by taking an appropriate limit. This is quite a 
strong constraint
on the polynomials, but not apparently strong enough to determine them.


\section*{The large-graph limit}

In statistical mechanics we are usually interested in the ``thermodynamic limit'',
when the system becomes large in such a way that bulk properties, such as the total free energy
or total entropy become proportional to the size of the system, and local or average properties,
such as the density, tend to a limit. 

For any regular lattice the ratio of the number of faces to the number of vertices
tends to a limit, determined by  the valency. For the honeycomb, square and
triangular lattices this ratio is $\half$, 1, 2, respectively. For any self-dual map
it is one. This suggest that we should consider the limit when $i$, $j$ become large, their 
ratio remaining fixed, non-zero and finite. This can be done in a symmetrical way by
defining
\be k = i+j \sep i = k \alpha \sep j = k \beta \comma \ee
and holding $\alpha, \beta$ fixed while $k \rightarrow \infty$. Note that
$\alpha + \beta = 1$.

In all the cases discussed above we find that the coefficients \\
$N(m,n,i,j)$, $N'(n,i,j)$, $P(i,j)$, $R(k)$ are asymptotically of the form
\be \label{asympt}
{\rm constant} \; \times \; \Kay^k /k^c \comma \ee
i.e. their ratio to $\Kay^k /k^c$ tends to a non-zero limit as $k \rightarrow \infty$.
This $\Kay$ is the ``partition function per site'' and is in general a function of 
$\alpha, \beta, x, y$; $c$ is an exponent, which for all our cases is a simple rational number.

Define 
\ba \lambda & = & (1 +  \alpha )^2 ( 1 +  \beta)  /( 4 \, \alpha ^3 ) \comma
\nonumber \\
\mu & = &  (1 + \alpha ) ( 1+ \beta)^2  / ( 4 \, \beta ^3 ) \period \ea
Then the results for $\Kay$ and $c$ are summarised in Table 1.

\begin{table} ~~~~~~~~~~~~~~~~~~~~~~~
{\renewcommand{\arraystretch}{1.3} 
\begin{tabular}{|cccc|}
\hline
~~~Case & Coeffs & $\Kay$ & $c$ \\ \hline
~~~1    &  $P$ & $x^{\beta} y^{\alpha}$ & -1  \\ 
~~~1   & $R$  &   max($x,y$) & 0 \\
~~~2   & $N, N', P$  & $\alpha^{-3 \alpha} \beta^{-3 \beta }$ & 9/2 \\
~~~2 & $R$  & 8 & 4   \\
~~~3 & $N, N', P$ & $\lambda^{\alpha} \mu^{\beta}$ & 3 \\
~~~3 & $R$ & 27/4 & 5/2 \\
 \hline
\end{tabular} 
\caption{{\protect{\footnotesize Values of $\Kay$ and $c$ for the coefficients 
$N, N', P, R$ of
the three solved cases.}}}
}
\end{table}

For both cases 2 and 3, $\Kay$ and $c$ are the same for the coefficients
$N, N', P$, provided we keep $m,n$ fixed (or at least small compared with $k$) while we take the 
limit $k \rightarrow \infty$.
Thus $m,n $ affect only the proportionality factor in (\ref{asympt}): from the viewpoint 
of statistical mechanics
they are ``boundary effects'' corresponding to various weightings of the rootings.
Indeed, it seems that the extra weights we have introduced by rooting the 
maps are also such a boundary effect: our result for $R$ in case 3 can be compared
with  eqn. (3.21) of Zinn-Justin and Zuber, obtained by matrix model 
techniques.\cite{ZinnJustinZuber99} Allowing for 
an extra integration, the two results agree to
within a constant factor (independent of $k$).

It is not actually necessary to calculate the coefficients to obtain $\Kay$.
From (\ref{defPR}) and (\ref{asympt}), the radius of convergence of the series for
$\Phi'(1,1,v,v,x,y)$ in powers of $v$ is $1/\Kay$, so $1/\Kay$ is the position of the 
closest
singularity to the origin in the complex $v$-plane of $\Phi'(1,1,v,v,x,y)$. For instance, 
for case 3 this is where the relation (\ref{v1v2}), with $v_1 = v_2$, $r = p$, i.e. 
$v = p(1-p)^2$ ceases to be invertible. This is when $dv/dp = 0$, i.e. $p = 1/3$,
$v = 4/27$, so $\Kay = 27/4$, as in Table 1.

Similarly, if $v_1 \neq v_2$ and $p \neq r$, then the singularity is when the Jacobian
$\partial(v_1, v_2 )/\partial (p, r)$ vanishes, which is when $3 p r + p + r = 1$, i.e. 
when $r = (1-p)/(1+3 p)$. Thus 
\be v_1^i v_2^j \eq \left(\frac{16 p^3}{ (1+3p)^2} \right)^{k \alpha} 
\left( \frac{(1-p)^3}{1+3p } \right) ^{k \beta } \period \ee
For given $\alpha,\beta$, this is maximized when $p$ is chosen so that
\be p \eq \alpha /(\alpha+ 2 \beta ) \sep r = \beta /(2 \alpha +  \beta ) \comma \ee
so then
\be 
v_1^i v_2^j \eq  \lambda^{- k \alpha} \mu^{- k \beta} \comma \ee
where $\lambda, \mu$ are defined as above. From Table 1, this is 
$\Kay^{-k}$, so the double series (\ref{defPR}) for $\Phi'(1,1)$ then just fails to converge,
as of course it should.

Further discussion of the techniques available to handle the asymptotic behaviour in case 3
is given by Bender and Richmond.\cite{Bender86}

\section*{A conjectured extension of case 3: $y = 1$ and $x$ large}

Ultimately it seems we should try to expand about case 3.
We noted in that section that there are actually three sub-cases that reduce to 
the enumeration problem,
namely $x+y = xy$, $x$ large and $y$ arbitrary, and $y$ large and $x$ arbitrary. 
We can consider an expansion about the second sub-case, expanding in inverse powers 
of $x$ while keeping $y$ fixed. If we choose $y = 1$, then we have observed a 
pattern in the coefficients 
$P(i,j)$, namely that to second-next-to-leading order the RHS of (\ref{Pij3}) becomes
\bd
4  x^j \frac{ 4 \,  (i+2 j-2)! \, (2 i +j -2)! }{
(i-1)! \, (2 i)! \, (j-1)! \, (2 j)! } \left( 1 + \frac{i}{x} + 
\frac{i (i+1) + 2 h}{ 2 \, x^2} + O(x^{-3}) \right) \comma \ed
where \bd h \eq \frac{i(i-1)(2 i-1)(i+j) (17-9i +3 j-3ij -2 j^2)}{
 (j+1) (2 j+1) (2i+j-2) (2i +j-3)( 2i+j-4) } \period \ed
We conjecture that this is correct for all $i, j \geq 1, 2i+j >4 $. It does have 
the expected behaviour
(\ref{asympt}) in the asymptotic limit. The partition function per site $\Kay$ is modified, 
acquiring an extra factor  
\bd
1 + \frac{\alpha}{x} + \frac{\alpha+\alpha^2}{2 \,x^2} - 
\frac{\alpha^3 (\alpha+2)}{\beta (\alpha + 1)^3 \, x^2 } + \cdots  \ed 
(remember that $\alpha + \beta = 1$). To this order the exponent $c$ remains 
at $c = 3$: there are no 
logarithmic terms in $k$ that would imply
a variation in this exponent.\cite{KadanoffWegner71}

\section*{Summary}

The functional relations developed by Tutte \cite{Tutte71} and Liu \cite{Liu90} 
give equations that define the sum of dichromatic poynomials over
rooted non-separable planar maps. This is the partition function of the Potts model,
summed (i.e. randomized) over such maps. 

We are interested in the partial sum over maps with a given number of sites and vertices 
($i+1$ and $j+1$, respectively), or with a given number $k = i+j$ of edges.
We are particularly interested in the ``thermodynamic'' limit when $i, j, k$ all 
become large, the ratios $\alpha = i/k, \beta = j/k$ remaining finite. We then 
expect the partial sums to behave 
asymptotically as (\ref{asympt}). The quantity  $\Kay$ therein is the ``partition function 
per edge''. We expect it to depend on $x, y, \alpha, \beta$ (or just $x, y$ if it is derived 
from $R(k)$), but not on the parameters 
$m, n$ which relate to the way in which the maps are rooted. For this reason we would be 
happy to restrict attention to the case  $u_1 = u_2 = 1$ (or any other fixed values of $u_1, u_2$), 
but we need to keep them arbitrary for the functional relations (\ref{fnreln})
as written to define the sum $\Phi$.

For the regular square lattice we know that the limiting function
$\Kay(x,y)$ has a singularity at $x = y$. For $x > 3$ (i.e. $q >4 $) this takes the form
of a discontinuity in  $\partial \Kay /\partial x$ as one crosses the line $x = y$.
For $1 < x \leq 3$  ($0 < q \leq 4$) there is no discontinuity, but there is a non-analyticity.
Physically this means that the Potts model has a phase transition at the self-dual point
$x = y$, which is
first order for $q >4$, continuous for $1 < q \leq 4$.\cite{Baxter73,Wu82}

The same is true for the triangular and honeycomb lattices, except that because these 
lattices are not self-dual, the transition is now at the 
point which maps to itself under a duality plus a star-triangle transformation. For 
the triangular (honeycomb) lattice this is when $y^2 + y = x+1$ 
($x^2 + x = y + 1$).\cite{BaxTempAsh78}

The obvious question is whether the $\Kay$ of this paper has similar properties, 
or whether the transition is removed by the 
randomization introduced by summing over maps. For the regular lattices one can develop
expansions for large $x$ and $y$ that make it quite clear that there is a first order 
transition. Unfortunately this technique fails for our randomization. From Table 1 
the exponent $c$ is different for the three solved cases that we have presented. There are two explanations for
this: either it varies continuously from case to case, or it varies discontinuously.
We conjecture that the latter is true: all the values presented are integers or half-integers, and 
it seems likely that this is always true. Cases 1 and 2 are the large and small $x,y$ limits, 
and it is likely that one is not allowed to interchange these limits with the large-$k$ 
asymptotic limit. This suggests that $c$ has the values of case 3 (namely $3$ and $5/2$) for all
positive finite $x$ and $y$, and this agrees with the second-order expansion of
the previous section.

So to investigate the phase transition problem we should focus on finite positive $x, y$
and look to see if $\Kay (x,y,\alpha,\beta)$ has non-analyticities for positive 
$x,y,\alpha,\beta$, in particular whether it is singular  across the self-dual line $x = y$
when $\alpha = \beta$.

In one sense the problem is solved: one can iteratively solve (\ref{fnreln}) in increasing 
powers of $v_1$ and $v_2$, the computing needed growing polynomially with the number of 
terms. A.J. Guttmann and others have argued that this is effectively a solution: it is 
exponential growth
(which is usual in statistical mechanical problems) that can make a problem intractable numerically.

Of course the problem with this is that one needs to take the limit of high powers of 
$v_1$ and $v_2$ in order to 
evaluate $\Kay$. Any finite powers will only give a numerical approximation to $\Kay$, and it 
is notoriously difficult to determine non-analyticities from numerical approximations. One could
focus on the case $v_1 = v_2 = v$, thereby calculating $R(k)$, and assign numerical 
values to $x$ and $y$. This would still leave one with the problem of handling a series expansion 
in the 
three variables $u_1, u_2, v$, so the calculation of $R(k)$ would grow with $k$ as $k^3$. It may be 
possible to make progress in this way, but it would certainly be a major numerical exercise.

One could reduce the problem to one of working with functions of a single variable
$u_1$ or  $u_2$ (but not both) by generalizing the arguments given after 
(\ref{Qeqn}), which would reduce the memory problems in a numerical calculation. It would of 
course be even better to remove both $u_1$ and $u_2$, or even obtain explicit equations
for the asymptotic 
behaviour in the limit $k \rightarrow \infty$. This can be done in case 3: 
can it be done in general?

Since writing these last paragraphs we have numerically investigated the sum of $P(i,k-i)$ 
in (\ref{PRsum}). If $x = y$, then $P(i,j) = P(j,i)$, so the terms in the sum are symmetrical about $i = k/2$.
For sufficiently small $x$ there is a single maximum, at $i = k/2$, and this dominates the 
sum in the limit of $k$ large. For sufficiently large $x$ there are two maxima, symmetric about 
$i = k/2$, which together dominate the sum.   If $x$ is slightly different from $y$,
 one will 
be larger than the other. As $k$ increases, both will grow exponentially, and the larger
will by itself  dominate the sum. Hence one can define an order parameter:
\be
M(x) \eq \lim_{x \rightarrow y^{+}} \lim_{k\rightarrow \infty} 
\sum_{i = 1}^{i=k-1} (k-2i) P(i,k-i)/[k R(k)] \period \ee
This will be zero for $x \leq x_c$, non-zero for $x > x_c$, where $x_c$ is the value of 
$x$ at which $P(i,k-i)$  first becomes two-peaked. Numerical studies 
up to $k = 56$ are reported in Appendix D. They  indicate 
that $x_c \simeq 9.5$, corresponding to $q_c \simeq 72$. Above this value there 
is a discontinuity in the first derivative of the limiting value $\Kay$ of 
$ R(k)^{1/k}$ 
across the line $x = y$ in the $(x,y)$ plane. This is a first-order transition.

For the regular lattices, the singularity in the free energy of the self-dual 
Potts model at $q = q_c = 4$ is essential (going like the exponential
 of $-1/\sqrt |q-q_c|$ ) : all derivatives exist and are continuous,
but the function is non-analytic \cite{Baxter73,Baxter82}. 
It would be interesting to determine if the same is true 
for this random model.

\section*{Acknowledgements}
The author belatedly thanks Professor W.T. Tutte for introducing him to 
this problem, and for showing how the result (\ref{conj}) can be proved 
recursively. He is indebted to Professor P. Zinn-Justin for correspondence regarding 
the correspondence between Tutte's approach and that of matrix models, in particular  
the asymptotic equivalence of the above results for case 3 with those of 
ref. \cite{ZinnJustinZuber99}.  The author also 
thanks V.V. Bazhanov for showing that the 
the formula for $p$ as a function of $v$ can be obtained from \cite{Bateman53}, and 
A.J. Guttmann for observations on what constitutes a solution in 
statistical mechanical problems.

\section*{Appendix A}
\setcounter{equation}{0}
\renewcommand{\theequation}{A\arabic{equation}}

Here we give the derivation of Liu's functional relation (\ref{fnreln}).
If $A$ is any edge of a graph ${\cal G}$ that is not a loop or an isthmus, let ${\cal G}'_A$ 
be the graph obtained from ${\cal G}$ by deleting the edge $A$, and ${\cal G}^{''}_A$ 
the graph obtained by identifying the two ends to form a single vertex. Then 
Tutte \cite{Tutte71} gives the following theorem:

\vspace{0.7cm}

THEOREM 1.  

\be \chi({\cal G};x,y) \eq \chi({\cal G}'_A;x,y)  + \chi({\cal G}^{''}_A;x,y) \period \ee

\vspace{0.5cm}

Also, if $\cal G$ is the union of two subgraphs ${\cal H}_1$ and ${\cal H}_2$ having 
just one vertex in common (so $\cal G$ is separable), then Tutte also gives

\vspace{0.7cm}

THEOREM 2.  

\be \chi({\cal G};x,y) \eq \chi({\cal H}_1;x,y) \chi({\cal H}_2;x,y) \period \ee

\vspace{0.5cm}

Tutte used these theorems to obtain a functional relation for the generating function
of the sum of dichromatic polynomials over all rooted maps. Here we shall adapt his method 
to the sum over all non-separable rooted maps.

The only non-separable maps with a loop or an isthmus are the first 
two graphs in Figure \ref{simple_graphs}. Applying Theorem 1 to all other maps, we obtain
\be \label{phicalc1}
\Phi \eq  x u_1^2 u_2 v_2 + y u_1 u_2^2 v_1 + Z_1 + Z_2 \comma \ee
where 
\ba \label{Z1Z2}
Z_1 & \eq & \sum \chi(M'_A; x,y) u_1^m u_2^n v_1^i v_2^j \comma \nonumber \\
Z_2 & \eq & \sum \chi(M^{''}_A; x,y) u_1^m u_2^n v_1^i v_2^j \comma \ea
the sums being over distinct non-separable rooted maps $M$ with at least two edges.

\begin{figure}[hbt]                    
\centering
\includegraphics[scale=0.7,angle=0]{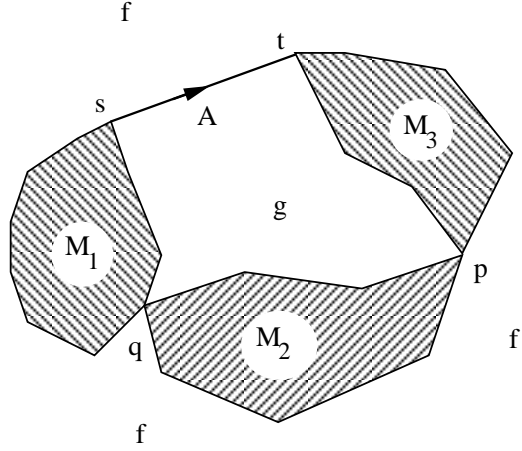}     
\caption{\protect\small A map $M$ that separates after 
removal of the root edge $A$. Each shaded region represents a non-separable
map with an outer face of at least two edges.}
\label{nodes}
\end{figure}

The map $M'_A$ may be separable, but only in a very specific way: if $s$ is the root vertex of $M$
and $t$ the other end of the edge $A$, there may 
be a sequence of nodes through which any route (after deletion of the edge $A$)
from $s$ to $t$ must pass. These, and only these, are vertices at which $M'_A$ can be separated. For instance, in Figure \ref{nodes} all routes from $s$ to $t$
(not along $A$) must pass sequentially through the two nodes $p$ and $q$. 
If $f$ is the root face of $M$, and $g$ the
other face adjacent to $A$, then $p$ and $q$ are points where the faces $f$ and $g$ touch.


\vspace{2.5mm}                                     
 
Consider the situation shown in Figure \ref{nodes}, where  $M'_A$ consist 
of three non-separable maps $M_1, M_2, M_3$ linked at $p$ and $q$. We ask how many
ways three such given rooted maps can be combined as in the Figure. For $M_1$ we can 
take the root-vertex to be $s$, the root-face to be $g$, and the root-edge to be
the first edge from $s$ alongside $g$. Similarly, $M_2$ has root-vertex $p$ and root-face  $g$;
$M_3$ has $q$ and $g$. Let $m_{\alpha}, n_{\alpha}, i_{\alpha}, j_{\alpha}$ be the values of $m,n,i,
j$ for $M_{\alpha}$. Also, let the number of edges of $M_{\alpha}$ adjacent to the face $g$ be
$r_{\alpha}$. Then $m_{\alpha} > r_{\alpha} \geq 1$, $n_{\alpha} >1$. Also,
 $m = m_1 + m_2 + m_3 - r_1 - r_2 - r_3 + 1$, 
$n = n_1 + 1$, $i = i_1 + i_2 + i_3 + 1$, $j = j_1 + j_2 + j_3$.
Using Theorem 2, the total contribution to $Z_1$ from the terms indicated by Figure \ref{nodes} 
is therefore
\bd
u_1 u_2 v_1 \; \sum_{M_1} \sum_{r_1 = 1}^{m_1 - 1} u_1^{m_1-r_1}  u_2^{n_1} v_1^{i_1}
v_2^{j_1} \chi(M_1;x,y) \ed
\be \label{sums}
\sum_{M_2} \sum_{r_2 = 1}^{m_2 - 1} u_1^{m_2-r_2}  v_1^{i_2}
v_2^{j_2} \chi(M_2;x,y) \; \; \sum_{M_3} \sum_{r_3 = 1}^{m_3 - 1} u_1^{m_3-r_3}  v_1^{i_3}
v_2^{j_3} \chi(M_3;x,y) \period \ee
Note that $u_2$ occurs only in the outer factor and the first sum.

Strictly, the $M_{\alpha}$-sums should be restricted to non-separable maps $M_{\alpha}$ 
with $m_{\alpha} \geq 2$.
However, the only 
non-separable map with $m <2$ is the second map (the loop-map) in Figure \ref{simple_graphs}, with
$m = 1$. If $m_{\alpha} = 1$, there are then no terms in the $r_{\alpha}$ summation, so 
the loop-map does not contribute and we can extend the summations 
to all non-separable maps.

Each $r_{\alpha}$ summation over $ u_1^{m_{\alpha}-r_{\alpha}}$ gives a factor
\bd 
\frac{u_1 - u_1^{m_{\alpha}}}{1 - u_1 } \period \ed
(Note that this vanishes for $m_{\alpha} = 1$.)
Inserting these, using (\ref{defnPhi}) and writing $\Phi(u_1,u_2,v_1,v_2,x,y)$ simply as 
$\Phi(u_1,u_2)$,  (\ref{sums}) reduces to
\be \label{sums2}
u_1 u_2 v_1 C(u_1,u_2)  C(u_1,1)^2 \comma \ee
 where $C(u_1, u_2) $ is defined by (\ref{defC}).

Figures like Figure \ref{nodes}, but with $k + 1$ separable pieces instead of three, give 
the same contribution (\ref{sums2}), but with $C(u_1,1)^2$ replaced by $C(u_1,1)^k$.
Summing over $k= 0,1,2,\ldots$, we obtain
\be
Z_1 \eq u_1 u_2 v_1 C(u_1,u_2) /[1-C(u_1,1)] \period \ee

To calculate $Z_2$ we note that $Z_2$ is the dual of $Z_1$ and obtain immediately
\be
Z_2 \eq u_1 u_2 v_2 D(u_1, u_2) / [1-D(1,u_2)]  \comma \ee
where $D(u_1, u_2) $ is also defined by (\ref{defC}).

From (\ref{phicalc1}), we therefore obtain the functional relation (\ref{fnreln}). 
This is equation (4.17) of \cite{Liu90}, $\mu, \nu, x, y, z, t$ therein being
our $x, y, v_2, v_1, u_1, u_2$.

\section*{Appendix B}
\setcounter{equation}{0}
\renewcommand{\theequation}{B\arabic{equation}}

Let $R (k)$ be the number of rooted non-separable maps with $k$ edges, and 
$U (k)$ the number of unrooted ones. Then $R(1), R(2), \ldots = $
2,1,2,6, 22, 91, 408, 1938, 9614, 49335,... and  $U(1), U(2), \ldots = $
2,1,2,3,6, 16, 42, 151, 596, 2605,... 

Also let $\phi (k)$ be the Euler totient function,
i.e. the number of integers less than $k$ that are relatively prime to $k$ (including 1). Thus
$\phi(1), \phi (2) , \ldots = $ 1,1,2,2,4,2,...

Then Liskovets and Walsh \cite{Liskovets83}
show that 
\be \label{LWreln}
4 k U (k) \eq 2 R (k) + \sum_{j|k} \phi(k/j) ( 9 j^2-9 j+2)
  R (j) + k \,  c_k \, R(k') \comma \ee
where
\ba c_k  = (k+1), & & k' = (k+1)/2 \; \; \; {\rm if \; \; }k
{\rm \; \; is \; \; odd} \comma \nonumber \\ 
c_k  = (3k-4)/4, & & k' = k/2 \; \; \; {\rm if \; \; }k
{\rm \; \; is \; \; even} \comma \ea 
the sum being over all positive integers $j$ that divide $k$ and are less than $k$.

Since this $R (k)$ is known, being given by (\ref{Rk}) with $x = 1$, the relation (\ref{LWreln})
defines $U(k)$. The number of ways of rooting a given map is at most twice the number of edges
(two directions for each edge), so
\bd R (k) \leq 2 k \, U (k) \period \ed
Numerically we observe that $2 k U(k)/R(k)$ decreases to unity exponentially fast,
suggesting that this upper bound on $R(k)$ is attained asymptotically.

\section*{Appendix C}
\setcounter{equation}{0}
\renewcommand{\theequation}{C\arabic{equation}}

To solve the equations (\ref{fnreln2}), set $u_2 = 1$ in the
equation not involving $\hat D$ and write
$t(u_1)$ for $T(u_1,1,v_1,v_2)$. We can eliminate the function $\hat C$. 
Writing $u_1$ simply as 
$u$, we obtain the functional relation
\be t(u) \eq [u v_1 + t(u)] \left[ u v_2 + \frac{u t(1) - t(u)}{1-u} \right] \period \ee
Writing $t(1)$ as $t_1$, this can be written as
\ba \label{quadtu}
t(u)^2 & + & (1-u+u^2 v_2-u v_2 + u v_1 - u t_1 ) t(u) + \nonumber \\
&&   u^3 v_1 v_2 -u^2 v_1 v_2 -u^2 v_1 t_1 \eq 0
\period \ea
Given $t_1$, this is a quadratic equation for $t(u)$. Its discriminant is a quartic in $u$.
We consider the situation when $v_1$ and $v_2$ are both small and of order $v$.
This quartic then has two zeros close to one, and two
of order $1/v$.

The solution $t(u)$ of the quadratic is certainly analytic for sufficiently small $u$, so 
has a convergent Taylor expansion. 
For a given map $M$ (other than the link map) of type $(m, n, i, j)$,
 $i \geq 1$ and $j \geq m-1$, so 
$i + j \geq m$. The coefficient of $u^m$ in the expansion of $t(u)$ is
therefore  no bigger 
than order $v^m$, so the radius of convergence is of order $1/v$. 
It follows that the two zeros of the
discriminant close to one must coincide, otherwise $t(u)$ would have a square root 
singularity  at them. 

This condition gives a fifth-degree equation for $t_1$ in terms of $v_1$ and $v_2$.
We do not write it down here as it is far better to proceed as follows. As a polynomial 
in $u$, the discriminant has the form $1 + \cdots + v_2^2 u^4 $. It must therefore
be identical to the polynomial $(1- e u)^2 (1+2 f u + s^2 u^2)$, where $e,f,s$ are parameters
to be determined.  Equating coefficients of $u,.\ldots , u^4$, we obtain four
relations between $v_1, v_2, t_1, e, f, s$, the last of which can be written as $v_2 = e s$.
If we then define $p, r$ so that
$e = 1-p, s/e = r$, we find $f = r (2 p r - p -1)$ and that $v_1, v_2, t_1 = T(1,1)$ 
are given by (\ref{v1v2}) and (\ref{T11}).

We thus have a parametrisation in terms of the variables $p, r$. Given $v_1, v_2$, 
we can in principle solve for $p, r$ and then obtain $T(1,1)$. We should choose the solution
where $p \simeq v_1, q \simeq v_2, T(1,1) \simeq v_1 v_2 $ when $v_1, v_2$ are both small. 
Note that $T(1,1)$
is plainly a symmetric function of $v_1, v_2$, in agreement with duality.


Writing $u$ again as $u_1$, we can go on to solve for $t(u_1)$ by replacing $u_1$ by 
$w_1$, where
\be u_1 = \frac{w_1 \, (1-p r w_1)}{1-r + r (1-p) w_1 } \period  \ee
As $w_1$ increases from 0 to 1, $u_1$ also increases from 0 to 1.
 
The  discriminant of the quadratic is now a perfect square and we find 
\be t(u_1) \eq T(u_1, 1) \eq \frac{p r \, w_1^2 \, [1 - p - r
 + p r^2 - p r (1-p) w_1 ]}
{1 -  r + r (1-p) w_1  } \comma \ee
choosing the solution which vanishes when $u_1$ and $w_1$ are zero.

If we also define $w_2$ in terms of $u_2$ by
\be u_2 = \frac{w_2 \, (1-p r w_2)}{1-p  + p (1-r) w_2 } \comma \ee
then from the duality symmetry
\be T(1,u_2) \eq \frac{p r  \, w_2^2 \, [1 - p - r + p^2 r - p r (1-r)w_2 ]}
{1 - p  + p (1-r) w_2  } \period \ee

It is now easy to solve (\ref{hatC}) - (\ref{fnreln2}) for the 
full function $T(u_1, u_2)$ (the equations are 
linear in this function), giving
\bd
T(u_1, u_2) \eq p r \, w_1^2 w_2^2 \, (1 - p r w_1) \, (1 - p r w_2) \; \times \ed
\be \label{tfull}
 \frac{(1-p)(1-r) - 
p r (1-p) w_1 - p r (1-r) w_2  }
{ ( 1- p r w_1 w_2 ) \; [1-r + r (1-p) w_1 ] \; [ 1-p + p (1-r) w_2] } \period \ee

This completes the parametric solution of the functional relation for $T(u_1,u_2)$, but it is still 
far from obvious that the coefficients in the expansion in powers of $v_1, v_2$ will
be anything straightforward. In fact they are (at least for $T(1,1)$ ~), as one can
observe empirically. The key to the proof 
is the identities (\ref{pmqn}), (\ref{pm}).

Let us look at the second (simpler) identity  (\ref{pm}) first. The rhs
is certainly some function of $m$: write it as $P (m)$. If (\ref{v1v2}) is true, then
$p$ is related to $v_1 = v_2 = v$ by $v = p ( 1-p)^2$. Multiplying by $p^m$, this implies 
the identity
\be \label{diffeqn}
P(m+3) -2 P (m+2) + P (m+1) - v P (m) = 0 \period \ee
This is easily verified from the series expansion (\ref{pm}) of $P (m)$.

But (\ref{diffeqn}) is a third-order difference equation. For integer $m$ it follows that
\be \label{pi(m)}
P  (m) =  A_1 p_1^m +  A_2 p_2^m +  A_3 p_3^m \comma \ee
where $p_1, p_2, p_3$ are the three roots of $v = p ( 1-p)^2$
and $A_1, A_2, A_3$ are some coefficients, independent of $m$. 

If we write (\ref{pi(m)}) down for three successive values of $m$,
i.e. $m, m+1, m+2$,
we can solve the resulting three equations for $A_1, A_2, A_3$
(provided $p_1, p_2, p_3$ are distinct, as they are for $|v| < 4/27$).
The result has the form
\be \label{A_j}
A_j \eq [c_{1 j} P(m) + c_{2 j} P(m+1) + c_{3 j} P(m+2) ]/p_j^m \comma \ee
where the coefficients $c_{1 j}, c_{2 j}, c_{3 j}$ are independent of $m$. 

For $v$ sufficiently small, we can bound the coefficients in the expansion
(\ref{pm}) and show that $P (m)$ is of order $v^m$.
The three roots of the cubic are close to $v, 1 ,1 $; we choose $p_1 = p \simeq v$,
$p_2 \simeq p_3 \simeq 1$. Now let $m \rightarrow \infty$  in (\ref{A_j}): the rhs
vanishes exponentially for $j = 2, 3$, so $A_2 = A_3 = 0$. Hence $P (m)$ is proportional 
to $p_1^m = p^m$. From (\ref{pm}),
$P (0) = 1$, so  $P (m) = p^m $ and we have proved (\ref{pm}) for sufficientl small
$v$. Both sides exist and are analytic for $|v| < 4/27$, so it is true throughout this domain.

When $n = 1$ the theorem implies that 
\be \label{p1}
p =  {\textstyle \frac{2}{3} } \left\{ 1 - F \left( \third, - \third ; \half ;
  {\textstyle \frac{27 v}{4}} \right) \right\} \comma \ee
a result that can be obtained
from equations (2.8.11) of \cite{Bateman53}. For all $n$,
\be P(n) \eq v^n \;  _{3} F_{2} \left( {\textstyle \frac{2 n}{3}},
{\textstyle \frac{2 n +1}{3}}, {\textstyle \frac{2 n+2}{3}} ;
n + \half, n+1 ; {\textstyle \frac{27 v}{4}} \right) \comma \ee
$_3 F _2$ being the generalized hypergeometric function (section 9.14.1 of Ref. \cite{GR}).

The first identity (\ref{pmqn}) can be proved in the same manner. One first eliminates $r$ 
between the equation (\ref{v1v2}) to obtain $
v_1 (1 - p)^4 = p[(1 - p)^2 - v_2]^2 $. This is fifth degree equation for $p$. When 
$v_1, v_2$ are small the roots are  approximately $v_1,1,1,1,1$.
Multiplying the equation by $p^m$ gives a fifth-order linear difference equation for the rhs
$P (m,n)$ of (\ref{pmqn}), as a function of $m$. We can verify that it is satisfied by 
the given series expansion.
Proceeding as before, it follows that
$P (m,n)/p^m$ is independent of $m$. By symmetry, $P (m,n)/(p^m r^n)$ is 
independent of both $m$ and $n$. We can verify that $P (0,0) = 1$, so $P(m,n) = p^m r^n$
and we have proved (\ref{pmqn}) for sufficiently small $v_1, v_2$. Both sides are 
analytic within the radius of convergence of the
double series, so the identity is true in that domain.

\section*{Appendix D}
\setcounter{equation}{0}
\renewcommand{\theequation}{D\arabic{equation}}

For $x = y$ and $k = 2 h$ an even integer, let $\gamma(k)$ be the relative 
difference between the central value 
(with $i = j = h$) of $P(i,j)$ and the adjacent value (with $i  = h-1$), so
$\gamma(k) = 1 - P(h-1,h+1)/P(h,h)$. This is 
a measure of the curvature of the function $P(i,k-i)$: positive for $x < x_c$, negative for
$x > x_c$. For $k$ finite we can define $x_c$ to be the value of $x$ for which
$\gamma(k)$ vanishes. As $k$ increases we expect it to tend to a limit, 
namely the bulk value of $x_c$ referred to above.

\begin{table} ~~~~~~~~~~~~~~~~~~~~~~~
{\renewcommand{\arraystretch}{1.3} 
\begin{tabular}{|rr|rr|rr|}
\hline
$k$   &    $x_c$ ~~~~~  & $k$   &    $x_c$ ~~~~~ & $k$   &    $x_c$ ~~~~~ \\  \hline
 4 & 11.52079729 & 22 & 9.13390695 & 40 & 9.48204802 \\
 6 & 9.10192128 & 24 & 9.21188431 & 42 & 9.49075809 \\
 8 & 8.57757720 & 26 & 9.27606631 & 44 & 9.49647477 \\
10 & 8.49125873 & 28 & 9.32844518 & 46 & 9.49963733 \\
12 & 8.55621629 & 30 & 9.37088248 & 48 & 9.50061434 \\
14 & 8.67512255 & 32 & 9.40499415 & 50 & 9.49971657 \\
16 & 8.80618487 & 34 & 9.43213746 & 52 & 9.49720720 \\
18 & 8.93047698 & 36 & 9.45343658 & 54 & 9.49331008 \\
20 & 9.04035623 & 38 & 9.46981849 & 56 & 9.48821637 \\
 \hline
\end{tabular} 
\caption{{\protect{\footnotesize The value of $x_c$ for $k = 4,6,\ldots ,56$.}}}
}
\end{table}

  We have calculated $x_c$ for $k=6,...,56$ and give the results in Table 2.
They do indeed appear to be converging to a limit rather less than 9.5, but 
it is hard to be more precise than this. One would expect such values 
to converge monotonically to a limit as an inverse non-integer power law in $k$, but 
these initially  decrease, then increase to a maximum at $k = 48$, 
and then start to decrease again. It looks as though
even higher values of $k$ are needed to examine the convergence. As a check of our 
numerical accuracy, we have performed the calculations to both 16-digit and 19-digit 
precision in Fortran (real*8 and real*12): the results agree to better than
the accuracy of Table 2.


\begin{thebibliography}{19}

\bibitem{Ambjorn95}
J. Ambj{\o}rn, G. Thorleifsson and M. Wexler, New critical 
phenomena in 2d quantum gravity, Nucl. Phys. B {\bf 439}(1995) 187 -- 204

\bibitem{Arques85} 
D. Arques, Une relation fonctionnelle nouvelle sur les cartes planaires 
point\'{e}es,
J. Comb. Theory B {\bf 39} (1985) 27 -- 42 

\bibitem{Arques99} 
D. Arques, Enumeration des cartes point\'{e}es sur une surface
orientable de genre quelconque en fonction des nombres de sommets
et de faces,
J. Comb. Theory B {\bf 77} (1999) 1 -- 24 

\bibitem{Bateman53}
H. Bateman and Erdelyi, Higher Transcendental Functions,
McGraw-Hill, New York, 1953

\bibitem{Baxter73}
 R.J. Baxter, Potts model at the critical temperature, J. Phys. C {\bf 6}
(1973) L445 -- L448

\bibitem{BaxTempAsh78}
 R.J. Baxter, H.N.V. Temperley and S.E. Ashley, Triangular Potts model at its 
transition temperature, and related models, Proc. Roy. Soc. {\bf A358} 
(1978) 535 -- 559 

\bibitem{Baxter82} R.J. Baxter, Exactly solved models in Statistical Mechanics,
Academic, London, 1982

\bibitem{Bender85}
E.A. Bender and N.C. Wormald, The number of loopless planar maps, 
Discr. Math. {\bf 54} 235 -- 237 (1985)

\bibitem{Bender86}
E.A. Bender and L.Bruce Richmond, A survey of the asymptotic behaviour of maps,
J. Comb. Theory B {\bf 40} 297 -- 329 (1986)

\bibitem{Brezin78}
E. Brezin, C. Itzykson, G. Parisi and J.B. Zuber, Planar diagrams, Commun. Math. Phys.
{\bf 59} (1978) 35 -- 51 

\bibitem{Brown63}
W.G. Brown, Enumeration of non-separable planar maps,
Can. J. Math. {\bf 15} 526 -- 545 (1963)

\bibitem{BrownTutte64}
W.G. Brown and W.T. Tutte, On the enumeration of rooted non-separable planar maps,
Can. J. Math. {\bf 16} 572 -- 577 (1964)

\bibitem{Daul95} 
J-M. Daul, $Q$-state Potts model on a random planar lattice,
hep-th/9502014 (1995)

\bibitem{DiFrancesco98}
P. Di Francesco, B. Eynard and E. Guitter, Coloring random triangulations,
Nucl. Phys. {\bf B516} (1998) 543 -- 587

\bibitem{Fortuin-Kasteleyn}
C.M. Fortuin and P.W. Kasteleyn, On the random-cluster model. I.
Introduction and relation to other models, Physica {\bf 57} (1972)
536 -- 564

\bibitem{GR}
I.S. Gradshteyn and I.M. Ryzhik, Tables of Integrals, Series and Products,
Academic, New York and London, 1965

\bibitem{Hooft74}
G. 't Hooft, A planar diagram theory for strong interactions, Nucl. Phys. 
{\bf B72} (1974) 461 -- 473

\bibitem{KadanoffWegner71}
L.P. Kadanoff, Some critical properties of the eight-vertex model, Phys. Rev.
B {\bf 4} (1971) 3989 -- 3993

\bibitem{Kasteleyn-Fortuin}
P.W. Kasteleyn and C.M. Fortuin, Phase Transitions in Lattice Systems with
Random Local Properties, J. Phys. Soc. Japan, Suppl. {\bf 26} 
(1969) 11 -- 14

\bibitem{Kostov95}
I.K. Kostov,  Solvable statistical models on a random lattice,
Nucl. Phys. B, Proc. Suppl. (Netherlands) {\bf 45A} (1996) 13 -- 28,
hep-th/9509124

\bibitem{Kostov00}
I. K. Kostov, 
Exact solution of the six-vertex model on a random lattice, 
Nuclear Physics B {\bf 575}  (2000)  513 -- 534

\bibitem{Liskovets83}
V.A. Liskovets and T.R.S. Walsh, The enumeration of non-isomorphic 
two-connected planar maps, Can. J. Math. {\bf 35} (1983)  417 -- 435

\bibitem{Liu88}
Y. Liu, Enumeration of rooted vertex non-separable planar maps, Chinese 
Annals of Mathematics {\bf 9} (1988) 390 -- 403

\bibitem{Liu90}
Y. Liu, On chromatic and dichromatic sum equations, Discr. Math. {\bf 84}
 (1990) 169 -- 179

\bibitem{Liu93}
Y. Liu, On functional equations arising from map enumerations, Discr. Math. {\bf 123}
 (1993) 93 -- 109

\bibitem{'t_Hooft74}
 G. 't Hooft, A planar diagram theory for strong interactions, 
Nucl. Phys. {\bf B72} (1974) 461 -- 473

\bibitem{Tutte63} W.T. Tutte, A census of planar maps, Can. J. Math. {\bf 15}
(1963) 249 -- 271

\bibitem{Tutte67} W.T. Tutte, On dichromatic polynomials, J. Comb. Theory {\bf 2}
(1967) 301 -- 320

\bibitem{Tutte68} W.T. Tutte, On the enumeration of planar maps, 
Bulletin Am. Math. Soc {\bf 74} (1968) 64 -- 74

\bibitem{Tutte70} W.T. Tutte, On Chromatic Polynomials and the Golden Ratio,
 J. Comb. Theory {\bf 9} (1970) 289 -- 296

\bibitem{Tutte71}
W.T. Tutte, Dichromatic sums for rooted planar maps,
Proceedings of Symposium in Pure Maths, Am. Math. Soc, 
{\bf 29} (1971) 235 -- 245

\bibitem{Tutte96} W.T. Tutte, Dichromatic sums revisited, J. Comb. Theory B {\bf 66}
(1996) 161 -- 167

\bibitem{Walsh75}
T.R.S. Walsh, Counting rooted maps by genus. III:Non-separable maps,
J.Comb.Theory B {\bf 18} (1975) 222 -- 259

\bibitem{Wexler93}
M. Wexler, Matrix models on large graphs,
Nucl. Phys. B {\bf 410} (1993) 377 -- 394

\bibitem{Whitney32} H. Whitney, The Colorings of Graphs, Ann. 
Math. (N.Y.) {\bf 33} (1932) 688 -- 718

\bibitem{Wormald81}
N.C. Wormald, On the number of planar maps, Can. J. Math. {\bf 33} (1981)
 1 -- 11

\bibitem{Wu82} 
F.Y. Wu, The Potts Model, Rev. Mod. Phys. {\bf 54} (1982) 235 -- 268

\bibitem{ZinnJustin99}
P. Zinn-Justin, The dilute Potts model on random surfaces,
Los Alamos Web Site: cond-mat/9903385

\bibitem{ZinnJustinZuber99}
P. Zinn-Justin and J-B. Zuber, Matrix Integrals and the Counting of Tangles and Links,
Los Alamos Web Site: math-ph/9904019

\end{thebibliography}
\end{document}